\documentstyle[12pt,amssymb,amscd]{article}
\newcommand{\nc}{\newcommand}

\nc{\dbar}{\bar{\partial}}
\nc{\be}{\begin{equation}}
\nc{\ee}{\end{equation}}

\catcode`\@=11

\@addtoreset{equation}{section}

\def\@normalsize{\@setsize\normalsize{15pt}\xiipt\@xiipt
\abovedisplayskip 14pt plus3pt minus3pt%
\belowdisplayskip \abovedisplayskip
\abovedisplayshortskip  \z@ plus3pt%
\belowdisplayshortskip  7pt plus3.5pt minus0pt}
\def\small{\@setsize\small{13.6pt}\xipt\@xipt
\abovedisplayskip 13pt plus3pt minus3pt%
\belowdisplayskip \abovedisplayskip
\abovedisplayshortskip  \z@ plus3pt%
\belowdisplayshortskip  7pt plus3.5pt minus0pt
\def\@listi{\parsep 4.5pt plus 2pt minus 1pt
            \itemsep \parsep
            \topsep 9pt plus 3pt minus 3pt}}

\def\underline#1{\relax\ifmmode\@@underline#1\else
        $\@@underline{\hbox{#1}}$\relax\fi}
\@twosidetrue
\relax

\catcode`@=12

\catcode`\@=11

\def\section{\@startsection{section}{1}{\z@}{3.5ex plus 1ex minus
   .2ex}{2.3ex plus .2ex}{\large\bf}}

\def\ps@headings{\def\@oddfoot{}\def\@evenfoot{}
\def\@oddhead{\hbox{}\hfill
        \makebox[.5\textwidth]{\raggedright\ignorespaces --\thepage{}--
        \hfill }}
\def\@evenhead{\@oddhead}
\def\subsectionmark##1{\markboth{##1}{}}
}

\ps@headings

\catcode`\@=12

\relax

\def\figcap{\section*{Figure Captions\markboth
        {FIGURECAPTIONS}{FIGURECAPTIONS}}\list
        {Fig. \arabic{enumi}:\hfill}{\settowidth\labelwidth{Fig. 999:}
        \leftmargin\labelwidth
        \advance\leftmargin\labelsep\usecounter{enumi}}}
 \relax
\def\tablecap{\section*{Table Captions\markboth
        {TABLECAPTIONS}{TABLECAPTIONS}}\list
        {Table \arabic{enumi}:\hfill}{\settowidth\labelwidth{Table 999:}
        \leftmargin\labelwidth
        \advance\leftmargin\labelsep\usecounter{enumi}}}
 \relax
\def\reflist{\section*{References\markboth
        {REFLIST}{REFLIST}}\list
        {[\arabic{enumi}]\hfill}{\settowidth\labelwidth{[999]}
        \leftmargin\labelwidth
        \advance\leftmargin\labelsep\usecounter{enumi}}}
 \relax

\catcode`\@=11

\def\ps@headings{\def\@oddfoot{}\def\@evenfoot{}
\def\@oddhead{\hbox{}\hfill
        \makebox[.5\textwidth]{\raggedright\ignorespaces --\thepage{}--
        \hfill }}
\def\@evenhead{\@oddhead}
\def\subsectionmark##1{\markboth{##1}{}}
}

\ps@headings

\relax

\def\firstpage#1#2#3#4#5#6{
\begin{document}

\begin{titlepage}
\nopagebreak
\title{\begin{flushright}
        \vspace*{-1.8in}
%        {\normalsize IC/01/???}\\[-9mm]
        {\normalsize hep-th/0112200}\\[4mm]
\end{flushright}
\vfill
{\large \bf #3}}
\author{\large #4 \\ #5}
\maketitle
\vskip -7mm
\nopagebreak
\begin{abstract}
{\noindent #6}
\end{abstract}
\vfill
\begin{flushleft}
\rule{16.1cm}{0.2mm}\\[-3mm]
$^{\star}${\small Research supported in part by the EEC contract \vspace{-4mm}
CT-2000-00148.}\\
December 2001
\end{flushleft}
\thispagestyle{empty}
\end{titlepage}}
\newcommand{\dal}{\raisebox{0.085cm}
{\fbox{\rule{0cm}{0.07cm}\,}}}
\newcommand{\dt}{\partial_{\langle T\rangle}}
\newcommand{\dtbar}{\partial_{\langle\bar{T}\rangle}}
\newcommand{\al}{\alpha^{\prime}}
\newcommand{\mst}{M_{\scriptscriptstyle \!S}}
\newcommand{\mpl}{M_{\scriptscriptstyle \!P}}
\newcommand{\dv}{\int{\rm d}^4x\sqrt{g}}
\newcommand{\lv}{\left\langle}
\newcommand{\rv}{\right\rangle}
\newcommand{\ph}{\varphi}
\newcommand{\sbar}{\,\bar{\! S}}
\newcommand{\xbar}{\,\bar{\! X}}
\newcommand{\fbar}{\,\bar{\! F}}
\newcommand{\zbar}{\,\bar{\! Z}}
\newcommand{\tbar}{\bar{T}}
\newcommand{\ubar}{\bar{U}}
\newcommand{\ybar}{\bar{Y}}
\newcommand{\phb}{\bar{\varphi}}
\newcommand{\cm}{Commun.\ Math.\ Phys.~}
\newcommand{\pr}{Phys.\ Rev.\ D~}
\newcommand{\pl}{Phys.\ Lett.\ B~}
\newcommand{\prl}{Phys.\ Rev.\ Lett.~ }
\newcommand{\ibar}{\bar{\imath}}
\newcommand{\jbar}{\bar{\jmath}}
\newcommand{\np}{Nucl.\ Phys.\ B~}
\newcommand{\e}{{\rm e}}
\newcommand{\gsi}{\,\raisebox{-0.13cm}{$\stackrel{\textstyle
>}{\textstyle\sim}$}\,}
\newcommand{\lsi}{\,\raisebox{-0.13cm}{$\stackrel{\textstyle
<}{\textstyle\sim}$}\,}
\date{}
\firstpage{95/XX}{3122} {\large\sc Little String
Theories in Heterotic Backgrounds$^{\star}$} {E.
Gava$^{a,b}$, K.S. Narain$^{ b}$ $\,$and$\,$
M.H. Sarmadi$^{\,b}$}%\\[-3mm]
%{\normalsize\sl
%$^a$Centre de Physique Th\'eorique, Ecole Polytechnique,$^\dagger$
%F-91128 Palaiseau, France\\[-3mm]
{\normalsize\sl
$^a$Istituto Nazionale di Fisica Nucleare, sez.\ di Trieste, SISSA,
Italy\\[-3mm]
\normalsize\sl $^b$The Abdus Salam International
Centre for Theoretical Physics,
I-34100 Trieste, Italy\\[-3mm]}
%\normalsize\sl $^d$Department of Physics, Northeastern
%University, Boston, MA 02115, U.S.A.}
{We study Little String Theories (LST) with ${\cal N}=(1,0)$ supersymmetry
arising, in a suitable double scaling limit,
from 5-branes in heterotic string theory or in
the heterotic-like type II/$(-)^{F_L}\times {\rm shift}$.
The limit in question, previously studied in the type II case,
is such that the resulting
holographically dual pairs, i.e. bulk string theory
and LST are at a finite effective coupling. In particular, the internal
$(2,2)$ SCFT on the
string theory side is non-singular and given by $SL(2)/U(1)
\times SU(2)/U(1)$ coset.
In the type II orbifold case, we determine
the orbifold action on the internal SCFT and construct the
boundary states describing the non-BPS massive states
of a completely broken $SO$ gauge theory, in agreement
with the dual picture of D5-branes in type II/$\Omega\times {\rm shift}$.
We also describe a different orbifold action which gives rise
to a $Sp$ gauge theory with $(1,1)$ supersymmetry.
In both the heterotic
$SO(32)$ and $E_8\times E_8$
cases, we determine the gauge bundles which correspond
to the above SCFT and
break down the gauge groups to
$SU(2)\times SO(28)$ and $E_7\times E_8$ respectively.
The double scaling limit in this case involves taking small instanton
together with small string coupling constant limit.
We determine the spectrum of
chiral gauge invariant operators with the corresponding global
symmetry charges on the LST side and compare with the massless
excitations on the string theory side, finding agreement for
multiplicities and global charges.}

\section{Introduction}
Little String Theories (LST) in type II or heterotic
backgrounds  received considerable attention in the
last few years\cite{seiberg, abks, gk1, gk2, gk, ds,
adg}. These 5+1 dimensional theories arise,
as first observed in \cite{seiberg}, by considering
the world-volume theory of 5-branes in the decoupling limit
where the string coupling constant $g_s$ goes to zero, in
such a way that the world-volume
theory decouples from the bulk string theory
but gives rise to a non-trivial, interacting 5+1 dimensional
Quantum Field Theory, with ${\cal N}=(2,0)$ or ${\cal N}= (1,1)$
supersymmetry
in type IIa or IIB, respectively. In both cases the theory contains
string-like excitations which however do not give rise to gravity.
In \cite{abks} it was
argued that type II string theory in the linear dilaton background
corresponding to a collection of coincident NS5-branes\cite{chs},
has a holographic description
in terms of the above 5+1 dimensional QFT. A class  of observables in the
LST, in short representations of the supersymmetry algebra,
was found to match with a class of excitations in the bulk string theory.
This analysis has been later extended in \cite{gk} to the heterotic string
in the background of symmetric 5-branes, which also give rise
to a SCFT involving a linear dilaton times an $SU(2)$ WZW model.
The problem with the linear dilaton
is that, no matter how small is the asymptotic value $g_s$ of the
string coupling, it gives rise to a divergent string coupling
near the NS5-branes, and therefore
is not amenable for perturbative analysis. It was further observed in
\cite{gk1,gk2} that, in the type II case,
this problem can be overcomed by separating the
NS5-branes: this has  the effect of cutting-off the strong coupling region
by adding a cosmological constant to the
corresponding Liouville CFT,or, equivalently, replacing the Liouville theory
with an $SL(2,{\bf R})/U(1)$ coset CFT. The full internal CFT
turns out to be an ${\cal N}=(2,2)$ SCFT, given
by the tensor product of a  ${\cal N}=(2,2)$ minimal
model times an ${\cal N}=(2,2)$ Liouville theory or $SL(2,{\bf R})/U(1)$
Kazama-Suzuki coset model
\footnote{Actually, in the internal,
world-sheet SCFT, supersymmetry is enhanced to ${\cal N}=(4,4)$. Earlier
work on (4,4) SCFT's related to
NS5-branes has appeared in \cite{kpr, afk, kounnas}}.
In this case one still takes
the decoupling limit $g_s\rightarrow 0$, but  holds fixed the tension
of the open D-string (in type IIB case) or open D2-branes
(in type IIA case) connecting the NS5-branes. On the LST side
this corresponds to moving off the origin of the Coulomb branch to
a point where the original $SU(2)_L\times SU(2)_R$ R-symmetry
group is broken down to $U(1)\times Z_N$, $N$
being the number of NS5-branes.
In \cite{gk1,gk2} various
tests of the holographic correspondence in this background
have been performed, including
the matching of short multiplets on the two sides.

Pourpose of this work is to extend the study of the holographic
connection to theories which have 8 supercharges in the background
of the 5-branes. In this case the LST has ${\cal N}=(1,0)$ supersymmetry.
Our main interest will be the heterotic
string, but we will consider also
a freely acting $Z_2$ orbifold of type II theory which removes
the spacetime supercharges coming from, say, the left-moving sector.
For this we require a compact $S^1$ direction and the $Z_2$ group element
will then be $(-)^{F_L}\sigma_P$, where $F_L$ is the left-moving spacetime
fermion number and $\sigma_P$ is a shift of order 2 along $S^1$.
The $S^1$ is considered to be part of the world-volume of the NS5-branes.
In this case, in a dual description of the system in terms of D5-branes
in type IIB/$\Omega \sigma_P$, it has been shown in \cite{ghmn1}
that one can have both $SO$ or $Sp$ gauge groups.
We will determine the appropriate action of the orbifold group on
the internal SCFT and construct boundary states. We will find that the
boundary states describe the massive gauge particles corresponding
to the generators of a completely
broken $SO$ gauge group, in agreement with the dual D5-brane picture.
As for
the $Sp$ case, we will find an orbifold action on the SCFT which results
into boundary states agreeing with an $Sp(N)$
group being broken down
to $Sp(1)^{N}$, with however unbroken ${\cal N}=(1,1)$
supersymmetry.
Thus the issue of reproducing a (1,0) $Sp$ gauge theory is still open.

As for the heterotic string case, we recall that in the analysis of
\cite{gk} the background corresponding to the linear dilaton (times $SU(2)$
WZW model) SCFT involved a singular gauge bundle with a single
fat $SU(2)$ instanton superimposed to $N-1$
small instantons \cite{witten, witten1} at the
origin of $R^4$. We will show that the SCFT involving the (2,2)
Liouville theory with cosmological constant (or equivalently the
$SL(2,{\bf R})/U(1)$ coset) corresponds to a non-singular gauge
bundle where the small instantons are given a (small) scale and
are separated in $R^4$. The double scaling limit
in this case involves taking small instanton
together with small string coupling limit.
On the LST side this corresponds to moving in the Higgs branch
to a point which has
a global symmetry group containing the (double cover of)
$U(1)\times Z_N$, as one expects from the world-sheet SCFT on
the string theory side.
We will compare, finding agreement, the
spectrum of a class chiral gauge invariant operators
from the LST, at the given point on the Higgs branch,
with the corresponding on-shell states from bulk string theory.

The paper is organized as follows: in section 2 we review the
main features of the (2,2) SCFT corresponding a configuration of 5-branes
in type II string theory, together with the construction of boundary
states describing the massive BPS states of the LST theory.
In section 3 we consider type II/$(-)^{F_L}\times\sigma_P$ in
the above background together with the (non BPS) boundary states
for the $SO$ case. We will also describe the  $Z_2$
orbifold giving rise to $Sp$ gauge group with (1,1) supersymmetry.
In section 4 we will turn to the heterotic $SO(32)$ and $E_8\times
E_8$ cases. First, we will describe the spectra of the two theories
in the mentioned SCFT, which corresponds to a symmetric embedding
of the gauge connection into the spin connection giving rise
to unbroken gauge group $SU(2)\times SO(28)$ and $E_7\times E_8$
in the two cases. We then analyze the ADHM constraints\cite{ADHM} for
$SO(4)$ instantons relevant for the $SO(32)$ theory, which
are nothing but the F- and D-term constraints for
a 4D, ${\cal N}=1$ gauge theory with $Sp(N)$ gauge group
and $SO(4)$ flavour group \cite{witten}.
We will identify the solution of
the ADHM constraints which determines the gauge bundle corresponding
to the SCFT under consideration and describe its global symmetries.
With the resulting massless spectrum we construct gauge invariant
chiral operators in the LST, compare with states one obtains from
string theory and verify the matching of the multiplicities and global
symmetry quantum numbers.

\section{Review of the Type II case}
In this section we will briefly review the CFT description of a
system of $N$ separated NS5-branes in type II string theories. Let
us remind that the CFT describing coincident NS5-branes \cite{chs}
has a bosonic part corresponding to the target space \be {\bf
R}^{5,1}\times {\bf R}_\varphi\times{ SU(2)}_{N-2}. \label{fi} \ee
The first factor describes the flat 5+1-dimensional Minkowski
part, corresponding to the world volume coordinates $X^\mu $ of
the NS5-branes. The second factor representing the radial
direction transverse to the branes, is actually described by a
Liouville field $\varphi$ with background charge $Q=\sqrt{2/N}$
(we set $\alpha'=2$ here). Finally, the third factor is given by
an $SU(2)$ WZW model at level $N-2$. Together with a total of 10
free fermions, the system has the critical central charge 15. In
\cite{abks} the above CFT has been used to test the holographic
correspondence between short multiplets in type IIA or IIB string
theory on this background and those expected to arise in the
corresponding 5+1-dimensional Little String Theories (LST).

However, string theory based on the above CFT is strongly coupled,
due to the linear dependence of the dilaton  on the coordinate
$\varphi$, which manifests itself with the background charge term
for $\varphi$. As a result, the string coupling constant diverges
for $\varphi\rightarrow -\infty$.

In \cite{gk1,gk2} it has been argued that separating the
NS5-branes, therefore moving in the Coulomb branch of the
5+1-dimensional world-volume theories, regularizes the strong
coupling singularity.

The Coulomb branch in the cases IIA nd IIB turns out to be
\cite{seiberg} $({\bf R }^4\times S^1)^N/S_N$ and $({\bf
R}^4)^N/S_N$ respectively, as dictated by the ${\cal N}=(2,0)$
respectively ${\cal N}=(1,1)$ nature of the 5+1 dimensional
supersymmetry. In both cases there are 4 scalars $X^i$,
$i=6,7,8,9$ in the adjoint of $U(N)$, parametrizing the transverse
positions of the NS5-branes, whose Cartan components give rise to
the above moduli spaces (in type IIA there is one more compact
scalar, giving rise to the additional $S^1$). Introducing the
complex scalars $A=X^8+iX^9$ and $B=X^6+iX^7$, one particular
point in the moduli space is characterized by the only non-trivial
gauge invariant vacuum expectation value
$<{\rm tr} B^N>=\mu^N$, corresponding to the $N$
fivebranes symmetrically distributed on a circle of radius $\mu$
in the (6,7) plane. This arrangement breaks the original $SO(4)$
transverse rotational symmetry (which is an R-symmetry of the
6-dimensional theory in type IIB case) down to $SO(2)\times Z_N$.
The double scaling limit is defined by taking $g_s\rightarrow 0$,
$\mu\rightarrow 0$, with $M_W=\mu/g_s \alpha'$ being held fixed.
This latter quantity is the ``W-boson'' mass, i.e. the mass of the
D-strings stretched between the NS fivebranes in the type IIB
case.

The regularization of the strong coupling singularity comes from
the fact that switching on a non-trivial vev for ${\rm tr}
B^N$ has the effect of adding a cosmological constant to the
world-sheet lagrangian for the Liouville field. To see this, is
convenient to rewrite the background (\ref{fi}) as: \be {\bf R
}^{5,1}\times {\bf R}_\varphi\times (S^1\times\frac{
SU(2)}{U(1)})/Z_N. \label{fi1} \ee where $S^1$ has radius
$\sqrt{2N}$ and its coordinate is denoted by $Y$. The
Kazama-Suzuki coset  $SU(2)/U(1)$ is an ${\cal N}=2$ minimal model
with central charge $3-6/N$,  and the $Z_N$ orbifold is
essentially the GSO projection. The field $\varphi+iY$, together
with its fermionic partners $\psi_L$, $\psi_R$ (and the conjugate
$\bar\psi$) makes an ${\cal N}=2$ superfield $\Phi$, giving rise to an
${\cal N}=2$ version of the Liouville theory, with central charge
$3+6/N$ \cite{es}. The perturbation corresponding to the
expectation value $<{\rm tr} B^N>$ can be
shown to be, in superfield notation

\be
\delta L=\int d^2\theta \exp{(-\frac{1}{Q}\Phi)}+c.c.
\label{pert}
\ee

The fact that the interaction in (\ref{pert}) grows exponentially
for $\varphi\rightarrow -\infty$ regularizes the strong coupling
singularity. One can show that the effective string coupling is of
order $1/M_W\sqrt{\alpha'}$.

A convenient dual (more precisely, mirror \cite{hk}) description
of the perturbed ${\cal N}=2$ Liouville theory, which we will use
in the following, is given by the non-compact $SL(2,{\bf R})/U(1)$
${\cal N}=2$ coset model \cite{vm, gv, ov, yamaguchi}. We are
thus equivalently lead to consider type II string theory on

\be {\bf R}^{5,1}\times (\frac{ SL(2,{\bf R})}{U(1)}\times\frac{
SU(2)}{U(1)})/Z_N. \label{fi2} \ee

In this picture the strong coupling region $\varphi\rightarrow
-\infty$ of the cylinder with coordinates ($\varphi$, $Y$) is cut
off and the infinite cylinder is replaced by a semi-infinite
cigar. The asymptotic radius of the cigar ($Q$) is related by
T-duality to that of the cylinder ($2/Q=\sqrt{2N}$).
The $SO(2)\times Z_N$ symmetry of the problem is manifest
in the perturbed Liouville CFT: the perturbation in (\ref{pert})
carries momentum N and zero winding along the circle parametrized
by $Y$, therefore winding is conserved (corresponding to
the $SO(2)$ symmetry), while momentum is conserved modulo
$N$ (corresponding to the $Z_N$ symmetry). Going to  the dual
description given by $\frac{ SL(2,{\bf R})}{U(1)}$, momentum
and winding are exchanged, in agreement with the cigar geometry
of the target space.

Finally, a useful correspondence comes from the fact that the CFT
$(\frac{ SL(2,{\bf R})}{U(1)}\times\frac{ SU(2)}{U(1)})/Z_N$ is
known \cite{ov} to describe (for the case of A-type modular invariants)
backgrounds given by (deformed) ALE spaces of the type $A_{N-1}$.
The defining equation in ${\bf C}^3$ of these non-compact
Calab-Yau twofolds is $x^N+y^2+z^2=\mu^N$, the
complex deformation parameter being $\mu^N$.
The precise statement is that type IIA(B) on ALE spaces of type
$A_{N-1}$ is equivalent to type IIB(A) string theory on the
background of N NS-fivebranes separated $Z_N$ symmetrically on a
circle \cite{ov, gk1, gk2}.

Since we can have arbitrarily small string coupling constant, we
can analyze the perturbative spectrum on the background
(\ref{fi2}) and compute scattering amplitudes among its states. To
do that we have to handle the non trivial, minimal model part of
the CFT $(\frac{ SL(2,{\bf R})}{U(1)}\times\frac{
SU(2)}{U(1)})/Z_N$.
Let us start from the coset $SU(2)/U(1)$, representing an ordinary
${\cal N}=2$ minimal model at level $N-2$: the left-moving part of
the primary fields is characterized  by three integers ${\ell},m,s$,
where $\ell=0,\dots, N-2$, $m=-N, \dots, N-1$ and $s=-1,0,1,2$, $0,2$
and $-1,1$ corresponding to the NS and R sector $SO(2)$ conjugacy
classes. There is a constraint ${\ell}+m+s={\rm even}$ and a $Z_2$
identification given by ${\ell}\rightarrow N-2-{\ell}$,  $m\rightarrow m+N$,
$s\rightarrow s+2$.
The corresponding conformal dimensions are given by
\be
\Delta=\frac{{\ell}({\ell}+2)-m^2}{4N}+\frac{s^2}{8},~~~~~~~
|m|\leq \ell ,
\label{delta}
\ee
Similarly, for the right-moving part we have (for A-type modular
invariants) the labels $\bar{\ell}={\ell}$, $\bar m$, $\bar s$. The
non-compact part $Sl(2,{\bf R})/U(1)$ is less understood, but we
will restrict to those ${\cal N}=2$ representations which
correspond to the $SL(2,{\bf R})$ discrete series with
${\ell}'=0,\dots,N-2$. We will have additional quantum numbers $m'$,
$s'$, ${\bar m}'$, ${\bar s}'$. $m'$ and ${\bar m}'$
take values in ${\bf Z}$ and are given
in terms of momentum $p$ and winding $w$
as $m'=p+wN$, ${\bar m}'=-p+wN$\footnote{Notice that here the
windings are multiple of $N$, however they become arbitrary
integers as aresult of the $Z_N$ orbifold.}.
The (left-moving) conformal
dimensions are given by:

\be \Delta'=\frac{m'^2-{\ell}'({\ell}'+2)}{4N}+\frac{s'^2}{8}, \label{delta'}
\ee and similarly for the right-moving ones.

An important fact for us is that
the minimal model has a $G=Z_N\times Z_2$ chiral symmetry group
which acts on the fields $\phi^l_{m,s}$ as
\be
\phi^{\ell}_{m,s}\rightarrow e^{2\pi i\frac{m}{N}} \phi^{\ell}_{m,s},~~
\phi^{\ell}_{m,s}\rightarrow (-)^s\phi^{\ell}_{m,s}.
\label{sym}
\ee
There is a similar
action for the fields of the non-compact $Sl(2,{\bf R})/U(1)$
part, with the formal flip of sign $N$ to $-N$, corresponding  to
a group $G'$ again equal to $Z_N\times Z_2$. The same
considerations apply to the right-moving sectors.

Physical states are constructed by tensoring states coming from
the spacetime part, the $Sl(2,{\bf R})/U(1)$ part and and
$SU(2)/U(1)$ part. As is familiar from Gepner's construction
involving ${\cal N}=2$ minimal models, one has to impose the GSO
projection condition on the total fermionic charge, which, in the
present situation, amounts to the following constraint:

\be
\frac{m-m'}{N}-\frac{s+s'}{2}-\frac{\vec{\delta}\cdot{\vec{w}}}{4}\in
2\bf{Z}+1, \label{gso} \ee
where $\vec{\delta}$ is the spinor
weight of $SO(4)$ and $\vec{w}$ is the $SO(4)$ weight of the
state. In addition, there are conditions ensuring that all
components are in the same sector (either R or NS):
\begin{eqnarray}
\frac{\vec{v}\cdot{\vec{w}}}{4}+\frac{s}{2} &\in & {\bf Z}\nonumber\\
\frac{\vec{v}\cdot{\vec{w}}}{4}+\frac{s'}{2} &\in & {\bf Z},
\label{same}
\end{eqnarray}
where $\vec v$ is the vector weight of $SO(4)$. Corresponding conditions
hold for the right-moving sector.

Notice that the effect of the $Z_N$ orbifold is to trivialize on
physical states the action of the diagonal
$Z_N$ subgroup of $G\times G'$, and that the
conditions (\ref{same}) identify the actions of the two $Z_2$'s.

The mass shell condition, in the NS sector and for primaries
with with $s=s'=0$, is given
by: \begin{eqnarray}
\frac{k_\mu
k^\mu}{2}+\frac{{\ell}({\ell}+2)-m^2}{4N}+\frac{m'^2-{\ell}'({\ell}'+2)}{4N}
&=&\frac{1}{2}~, \nonumber\\
\frac{k_\mu
k^\mu}{2}+\frac{{\ell}({\ell}+2)-{\bar m}^2}{4N}+
\frac{{\bar m}'^2-{\ell}'({\ell}'+2)}{4N}
&=&\frac{1}{2}~.
\label{mass}
\end{eqnarray}
Massless states (scalars) are
characterized by $m'=-\epsilon (l'+2)$, $m=\epsilon l$
and ${\bar m}'=-{\tilde\epsilon} (l'+2)$, $m={\tilde\epsilon} l$
with ${\ell}+{\ell}'=N-2$. The values $\epsilon, {\tilde\epsilon} =\pm 1 $
correspond, for a given $\ell$ $({\ell}')$, to the four (c,c), (c,a),
(a,c), (a,a) primary states of the internal ${\cal N}=(2,2)$ SCFT.

There are altogether $4(N-1)$ scalars,
matching  with the number of Cartan generators of $SU(N)$. Also their
$U(1)$ and $ Z_N$ charges, given by $m'-{\bar m}'$ and
$m'+{\bar m}'$ respectively, agree with those of the
gauge invariant chiral operators  ${\rm tr} A^{\frac{(m'-{\bar m}')}{2}}$
and ${\rm tr} B^{\frac{(m'+{\bar m}')}{2}}$ of the 5+1 world-volume theory.
The gauge fields (in the type IIB case say) appear in the R-R sector
and are obtained by spectral flow from the scalars in the NS-NS
sector. Their quantum numbers are $m={\ell}+1$, $m'=-{\ell}'+1$
and $s=s'=1$, with similar values in the right-moving sector.

Charged, massive states (``W-bosons'') should come as D-branes
(D-strings in the type IIB case) stretched between NS fivebranes,
and should have a boundary state representation within the CFT
discussed above. They are BPS states, preserving 8 of the 16
supercharges. The construction of boundary states for
Gepner-like\footnote{Open strings in minimal models and Gepner models have
been first discussed in \cite{sagnotti}}
models has been first performed in \cite{rs} and more recently
analyzed by several authors \cite{douglas, vafa}. One starts from
Ishibashi states, which are in one-to-one correspondence with
(diagonal) primary fields. One then constructs Cardy states, which
are linear combinations of Ishibashi states such that the
corresponding cylinder amplitudes admit the interpretation of
partition functions when rewritten in the open string channel.  In
our case, denoting Ishibashi states as $|l,m,s,l',m',s',\vec{w}>$
and Cardy states with the corresponding capital letters, we have,
up to an overall normalization : \be
|L,M,S,L',M',S',\vec{W}>=\sum_{l,m,s,l',m',s',\vec{w}}^{'}
S^{L}_{l}S^{L'}_{l'}e^{i\pi\frac{Mm-M'm'}{N}}
e^{i\pi(\frac{Ss+S's'}{2}+\frac{\vec{W}\cdot{w}}{2})}
|l,m,s,l',m',s',\vec{w}>, \label{cardy} \ee where \be
S^{L}_{l}=\frac{\sin\pi\frac{(L+1)(l+1)}{N}}{\sqrt{\sin\pi\frac{(l+1)}{N}}}
\label{modular} \ee (and similarly for $S^{L'}_{l'}$) is the
S-modular transformation matrix for the $SU(2)_{N-2}$ characters.
The sum in (\ref{cardy}) is actually restricted
to the states obeying the GSO condition plus the condition that
all their components belong to  the same sector, as given in
(\ref{gso}) and  (\ref{same}). Notice that (\ref{gso}) implies
that $m-m'$ is a multiple of $N$. These conditions can be imposed
in (\ref{cardy}) by starting with a free sum, introducing
multipliers $a, b, c= 0,1$, $r=0,\dots, N-1$ and then inserting in
the sum the projection factors: \be \frac{1}{N}
\sum_{r=0}^{N-1}e^{2\pi i\frac{(m-m')}{N}r} \frac{1}{2}
\sum_{a=0}^{1}e^{2\pi i(\frac{s}{2}+
\frac{\vec{v}\cdot\vec{w}}{4})a} \frac{1}{2} \sum_{b=0}^{1}e^{2\pi
i(\frac{s'}{2}+ \frac{\vec{\delta}\cdot\vec{w}}{4})b} \frac{1}{2}
\sum_{c=0}^{1}e^{\pi i(\frac{m-m'}{N}-\frac{s+s'}{2}-
\frac{\vec{\delta}\cdot\vec{w}}{4}+1)c} \label{mult} \ee This has
the effect of shifting the labels of the Cardy states, $M$, $S$.
$M'$, $S'$, $\vec{W}$, to \begin{eqnarray} \tilde{M}&=&M+2r+c,~
\tilde{M'}=M'+2r+c,~\tilde{S}=S+2a-c,~\tilde{S'}=S'+2b-c,\nonumber\\
\vec{\tilde{W}}&=&\vec W+(a+b)\vec v+c\frac{\vec\delta}{2}.
\label{shift} \end{eqnarray} Therefore we can choose orbit
representatives labelled by $L$, $M$, $S$ and $L'$. Furthermore we
set $L'=0$, as states with $L'>0$ correspond to multi D-brane states.

The number of states obtained this way is $N(N-1)$: using the
$Z_2$ identification mentioned before, we can choose $0\leq L \leq
\frac{N-2}{2}$, $0\leq M < 2N$ and $S=0,2$, with $L+M=\rm even$,
giving the expected number of roots of $SU(N)$.
One can proceed further to
prove that these states have the right charges, as in \cite{lerche,
lls, es}, by computing the intersection index, which gives the
Cartan matrix of $SU(N)$. Alternatively, one can evaluate directly the
charge vectors by taking the overlap of the Cardy states with the
R-R states corresponding to Cartan generators  of $SU(N)$, which
amounts to project to Ishibashi states with $m={\ell}+1$,
$m'=-{\ell}'+1$, $s=s'=1$ etc., as discussed before.
This gives for the charge vector
$q_l(L,M,S)$: \be q_l(L,M,S)=\sin \pi \frac{(L+1)(l+1)}{N}e^{-i\pi
\frac{M(l+1)}{N}} e^{i\pi\frac{S}{2}.} \label{charge} \ee By
taking inner products of these vectors one can check that they
span the root diagram of $SU(N)$. Notice that the actions of $Z_N$
and $Z_2$ on Ishibashi states, as given in (\ref{sym}),
induce on Cardy states the maps
$M\rightarrow M+2$ and $S\rightarrow S+2$ respectively. The latter
has the effect of flipping the sign of the charge vector $q_l\rightarrow
-q_l$, therefore it sends roots to their negatives, i.e. it reverses the
orientation of the D-branes.

Finally, the tension of the D-branes is computed by taking the
overlap of the Cardy states with the state corresponding to the
identity operator ($l=m=s=0$). One gets a factor proportional to
\be \sin\pi \frac{(L+1)}{N}, \label{tension}\ee which agrees
nicely with the geometrical picture of D-branes stretched between
$N$ NS-fivebranes distributed $Z_N$ symmetrically on a circle.

\section{The case of Type II/$(-)^{F_L}\sigma_P$}

We now concentrate our discussion on the case of NS-fivebranes in
a class of theories which, although considerably simpler, have
many feature in common with the heterotic string. These theories
are obtained by modding out type IIB string theory by
$(-)^{F_L}\sigma_P$ where $F_L$ is the left-moving spacetime
fermion number and $\sigma_P$ is a shift of half-period along a
circle $S^1$, with $P$ the corresponding momentum. In the
untwisted sector this has the effect of giving a phase $(-)^P$ to
the states with momentum $P$. As a result, the left-moving
spacetime supersymmetry is completely broken (twisted states are
massive). S-duality relates this theory to the orientifold of IIB
by $\Omega\sigma_P$, $\Omega$ being the world sheet parity
operator. The presence of the shift avoids the introduction of
D9-branes, allowing both $SO$ or $Sp$ projections for
D5-branes with longitudinal shift \cite{ghmn1}.
Various
non-perturbative aspects of these theories, including tests of
S-duality  have been discussed in \cite{bgmn, ghmn1, ghmn2},
whereas boundary states have been analyzed in \cite{gutp}.

We will consider the case where the shift $\sigma_P$ is
longitudinal to the fivebranes, say along the direction 5. The
background corresponding to $N$ separated NS-fivebranes should be
the same as the one in type IIB. So, string theory should  be
described by the CFT \be {\bf R} ^{4,1}\times {S^1}\times(\frac{
SL(2,{\bf R})}{U(1)}\times\frac{ SU(2)}{U(1)})/Z_N, \label{cft1}
\ee further modded by $(-)^{F_L}\sigma_P$.

Thinking in terms of the dual picture involving $N$ D5-branes in
type IIB/$\Omega \sigma_P$ with longitudinal shift, we have, to
begin with, a 5+1 dìmensional $SO(N)$ gauge theory  with ${\cal
N}=(1,0)$ supersymmetry (if $ N$ is even we can have also
$Sp(N/2)$ gauge group) with vector multiplets in the adjoint of
$SO(N)$ and hypermultiplets in the second rank, symmetric tensor
representation of $SO(N)$. The separation of the D5-branes, in the
symmetric configuration described in the previous section, is
achieved by giving vev to the hypermultiplets, and as a result the
gauge group is completely broken. The vector multiplets combine
with the off-diagonal hypermultiplets to give massive non-BPS
multiplets, leaving $N-1$ massless scalars (excluding
the center of mass degree of freedom).
They correspond to the Cartan generators of $SU(N)$, which,
from the $SO(N)$, ${\cal N}=(1,0)$ viewpoint, are the diagonal
components of the hypermultiplets in the symmetric tensor representation.
This is to be contrasted with the type IIB case, where
we had ${\cal N}=(1,1)$ supersymmetry and the "W-bosons" were
$\frac{1}{2}$-BPS states. In the ${\cal N}=(1,0)$ case the
supersymmetry algebra does not admit a central extension and
therefore there are no short massive multiplets. After compactifying
to 4+1 dimensions, a central term appears and therefore
there are short multiplets, which are
associated to the Coulomb branch. We are however interested in
the Higgs branch, since we give vev to the hypermultiplets, and
this produces long (i.e. 16 dimensional) massive multiplets.

Let us go back to type IIB/$(-)^{F_L}\sigma_P$ on the background
(\ref{cft1}) and first discuss the perturbative spectrum we
obtain: repeating the arguments of the previous section for the
type II case, one sees that in the NS-NS sector there are $N-1$
massless scalars,
as expected from the previous arguments.
In the notation introduced before they are
characterized by the same values of $m$, $\bar m$, $m'$,
${\bar m}'$, $\ell$, $\ell '$ given after equation (\ref{mass}).
However, due
to the projection by $(-)^{F_L}\sigma_P$, the R-R sector does not
give massless gauge fields, since R-R states with $P=0$ are
projected out. This is in agreement with the dual D5-brane picture
discussed in the previous paragraph, where we argued that the
gauge group is completely broken.

As for the non-BPS "W-bosons", corresponding to the generators of
$SO(N)$, they should again appear as (possibly unstable) non-BPS
D-branes, i.e. boundary states. The construction of Cardy states
proceeds similarly to the type II case, the only modification
arising from the introduction, in addition to (\ref{mult}), of
projection factors taking into account the $(-)^{F_L}\sigma_P$
projection. Consider first the action of $\sigma_P$: denoting
respectively by $|P>$ and $|x>$ the boundary states with given
momentum $P$, respectively position $x$ along the circle $S^1$
transverse to the D-brane, we will have
$|x>=\sum_P(-)^P e^{i\frac{P}{R}x}|P>$. Here $R$ is the radius
of $S^1$.
On the other hand, the operator $(-)^{F_L}$ can be
realized on Ishibashi states by $e^{i\pi s}$: from the
construction detailed in section 2, states with $s={\rm odd}$
belong to the left-moving R sector. Therefore, we insert
$\frac{1}{2}(1+e^{i\pi s})$ in (\ref{cardy}). This has the effect
of shifting the quantum number $S$ of Cardy states to $S+2$, i.e.
of sending a D-brane to an anti-D-brane. The boundary state
invariant under the $(-)^{F_L}\sigma_P$ projection is therefore a
superposition of a D-brane at $x$ and an anti-D-brane at $x+\pi
R$. It is easy to see that the
number of these states is $N(N-1)/2$, equal to the dimension of
$SO(N)$ as expected.

By computing, for instance, the cylinder amplitude involving the
above boundary state and then reading it in the open string
channel, one can analyze the open string excitations of the
D-brane /anti-D-brane system: the open string stretched between a
D-brane at $x$ and an anti-D-brane at $x+\pi R$ (or viceversa )
has a ground state whose mass is $R^2-\frac{1}{2}$, and therefore
it is tachyonic, i.e. unstable, for $R^2 < \frac{1}{2}$.

We have mentioned previously that for $N$ even (and with longitudinal shift)
the dual system of $N$ D5-branes admits
an $Sp(\frac{N}{2})$ ${\cal N}=1$
gauge theory, with vector multiplets in the symmetric
representation and hypermultiplets in the antisymmetric one. By
giving vev to the latter ones we can realize a configuration in
which $N/2$ pairs of D5-branes are distributed $Z_{\frac{N}{2}}$
symmetrically on a circle, thereby breaking the gauge group down
to $SU(2)^{\frac{N}{2}-1}$, together with  $\frac{N}{2}-1$
massless scalars (we neglect here too the center of mass
degrees of freedom). The question is how this configuration
is realized in the present case of $N$ NS-fivebranes i.e.
what is the corresponding CFT.

We have argued that the CFT  $\frac{ SL(2,{\bf
R})}{U(1)}\times\frac{ SU(2)}{U(1)}$, or the equivalent one
involving the ${\cal N}=2$ Liouville theory, describes the configuration
of $N$ fivebranes  distributed $Z_N$ symmetrically on $S^1$. As
mentioned in section 2, in terms of the ALE space description this
corresponds to the equation \be \prod_{k=0}^{N-1}(x-{\omega}^k
\mu)+y^2+z^2=x^N+y^2+z^2-\mu^N=0, \label{ale1} \ee with
$\omega=e^{\frac{2\pi i}{N}}$ the primtive $N$-th root of unity.
The deformation $\mu^N$ corresponding to the cosmological term
(\ref{pert}) in the ${\cal N}=2$ Liouville theory. The configuration
alluded to before, with $Z_{\frac{N}{2}}$ symmetry, should
correspond to \be \prod_{k=0}^{\frac{N}{2}-1}(x-{\omega}^{2k}
\mu)^2+y^2+z^2=0. \label{ale2} \ee Compared to the $Z_N$ case,
(\ref{ale1}), we see that there is an additional deformation,
corresponding to the monomial $x^{\frac{N}{2}}$.
It is easy to see that in turn this deformation
corresponds, in the $\frac{ SL(2,{\bf R})}{U(1)}\times\frac{
SU(2)}{U(1)}$ CFT, to the scalar with quantum numbers
${\ell}=\pm m=\frac{N}{2}$, $m'=\pm ({\ell}'+2)$,
with ${\ell}+{\ell}'=N-2$.

One can check, to first order, that indeed the above perturbation
has the correct effect on the D-brane tension. The first order
correction is basically given by the overlap of the boundary state
with the state corresponding to the scalar field (plus its
conjugate) and is proportional to \be \cos\frac{\pi
(L+1)}{N}[e^{i\pi\frac{M}{2}}+e^{-i\pi\frac{M}{2}}].
\label{perttens} \ee This expression also agrees with what one
would obtain geometrically in terms of branes wrapped
on 2-cycles of the ALE manifold. In this case  tensions are
given by the (modulus of) the holomorphic volumes of the homology
2-cycles, i.e the integrals of the holomorphic 2-form over the
homology 2-cycles. If one perturbs the original 2-form
corresponding to (\ref{ale1}) with $x^{\frac{N}{2}}$, then, to
first order one gets (\ref{perttens}).

It would be interesting to find the CFT to which the system is
driven by the above field. The problem cannot be addressed
perturbatively, since the strength of the perturbation is tuned to
the original cosmological term, as it is clear from (\ref{ale2}).
Moreover, the CFT in question is expected to be singular: the
space (\ref{ale2}) has vanishing cycles, therefore there are
tensionless D-branes, responsible for the enhancement of the gauge
symmetry to $SU(2)^{\frac{N}{2}-1}$. In any case, one may hope
that the resulting CFT has a symmetry $Z_{\frac{N}{2}}\times
(Z_2)^{\frac{N}{2}}$, and that modding out by a diagonal $Z_2$ produces
the expected spectrum.

If $N$ is even, one may consider gauging a different $Z_2$,
namely the diagonal subgroup of $Z_N\times Z_2$, whose action
on the minimal model fields is given by:
\be
\phi^l_{m,s}\rightarrow e^{i\pi (m+s)}\phi^l_{m,s},
\label{orb}
\ee
together with $\sigma_P=(-)^P$ on the $S^1$ part of the
tensor product fields.
Notice that, since $l+m+s={\rm even}$, $e^{i\pi (m+s)}=(-)^l$.
Supersymmetry is preserved by this orbifold projection, since
the supercharge generators have $m=s=1$, therefore we
have ${\cal N}=(1,1)$ supersymmetry in 5+1
dimensions.  Out of the $N-1$ massless
scalars found in section 2, those
with $l$ even survive the projection,
and this gives  $N/2$ states.
Accordingly, there are $N/2$ gauge fields from the RR sector.
The construction of Cardy states proceeds following the
general strategy indicated in section 2.
The action (\ref{orb}) on Ishibashi states induces the action
$S\rightarrow S+2$, $M\rightarrow M+N$ on Cardy states.
Therefore if we insert, in addition to (\ref{mult}),
the projector $\frac{1}{2}\sum_{d=0,1} \exp(i\pi(m+s+P)d)$ in (\ref{cardy}),
we will have, in the notation of section 2,   $\tilde{M}=M+2r+c+dN$.
We can therefore restrict $M$ to the range $0\leq M \leq N-1$.
Together with the allowed values $S=0,2$,
$0\leq L \leq \frac{N-2}{2}$, $L+M={\rm even}$, this
gives a total of $N^2/2$ boundary states, which should
correspond to the broken charged generators of the gauge
group. We have therefore,
including the $N/2$ Cartan generators, a group of rank $N/2$
and dimension $N(N+1)/2$. To decide whether it is $Sp(N/2)$
or $SO(N+1)$ we compute the lengths of the root vectors:
from (\ref{charge}) one can easily verify that
the (squared) length of a root vector depends from $L$ as:
\be
|\vec{q}(L,M,S)|^2=\sum_{l\, {\rm even}=0}^{N-2}
{\sin}^2 \pi \frac{(L+1)(l+1)}{N}=\frac{N}{2}(1+
\delta_{L,\frac{N-2}{2}}),~~L=0,\dots,\frac{N-2}{2}.
\label{length}
\ee
From (\ref{length}), taking into account the
allowed values of $M$ and $S$,
we see that there are $N$ long
and ${N^2}/{2}-N$ short roots, therefore
we conclude that the gauge group is  $Sp(N/2)$.

\section{Heterotic $SO(32)$ and $E_8\times E_8$ string}

We will now discuss the little string theory for the standard
heterotic string theory. The NS5 brane solution now involves,
besides the metric, antisymmetric tensor and the dilaton, also
the gauge fields. In the following we will restrict ourselves to
the symmetric 5-branes, in which the spin connection is identified
with the gauge connection. Since the spin connection sits in an
$SU(2)$ subgroup of the transverse $SO(4)$ group, the gauge field will
also be in an $SU(2)$ subgroup of $SO(32)$ or $E_8 \times E_8$.
The
unbroken gauge group is then $SO(28) \times SU(2)$ and $E_7 \times
E_8$
respectively. These latter groups therefore will play the role of
flavour symmetries in the 5-brane world volume theory.

\noindent{\bf CFT Analysis}

Let us first discuss the $SO(32)$ case.
The singular CFT for symmetric 5-brane background has been discussed
in  \cite{gk}. It is simply a left-right symmetric
$SU(2)$ super-WZWN model at level $N$ together with the super Liouville
field and the additional 28 free gauge fermions in the left moving
bosonic sector. One of the important point that was made in this paper
was that this CFT describes the situation where the $N$ 5-branes
charge appears in the form of $N-1$ coincident small scale instantons
and 1 large scale instanton. The world volume theory therefore only
sees $Sp(N-1)$ gauge group. The number of operators (e.g. those
charged under $SO(32)$) in this case will be proportional to $(N-1)$,
which is indeed the number of primaries of the $SU(2)$ WZW model
\footnote{A discrepancy pointed out in \cite{gk} between the string
theory states and the gauge theory operators involved the left moving
$SU(2)$ current algebra descendents of spin $(N-2)/2$ primary which
gives states transforming under spin $N/2$ of the global part of this
$SU(2)$.
The corresponding gauge theory operator was missing. However the point
is that such current algebra descendents also do not exist in the CFT
as they are null states in the Verma module.}.
This idea of having one large scale instanton, is therefore the
mechanism analogous to the factoring out of the center of mass $U(1)$
in the type II context.

What happens when we turn on the perturbation to go to the
non-singular CFT given by the product of the $SL(2)/U(1)$ times
$SU(2)/U(1)$ coset models. We will first obtain the results from the
CFT and then later identify the heterotic background that this CFT
corrseponds to. As mentioned earlier we will restrict ourselves to the
left-right symmetric case. What this means is that the internal theory
is left-right symmetric ${\cal{N}}=2$ minimal models based on
$SU(2)/U(1)
\times SL(2)/U(1)$. In fact in the present case ${\cal{N}}=2$ is
promoted to ${\cal{N}}=4$ theory. Besides this
of course we have the right moving
${\cal{N}}=1$ free superconformal theory of $R^4$ (in the light cone
gauge) and the left moving free bosonic theory of $R^4$ and 28
fermions. The Virasoro constraints now read:
\begin{eqnarray}
\frac{k_\mu
k^\mu}{2}+\frac{l(l+2)-m^2}{4N}+\frac{{m'}^2-\ell'(\ell'+2)}{4N} + \frac{s^2+{s'}^2}{8} + \Delta
&=& 1 \nonumber\\
\frac{k_\mu
k^\mu}{2}+\frac{\ell(\ell+2)-{\bar{m}}^2}{4N}+\frac{{\bar{m'}}^2-\ell'
(\ell'+2)}{4N}
+ \frac{{\bar{s}}^2+{\bar{s'}}^2}{8}+
{\bar{\Delta}}
&=& \frac{1}{2} ~~~,
\label{virasoro}
\end{eqnarray}
where $\bar{\Delta}$ is the contribution to $\bar{L}_0$ from the descendants of
the minimal models together with that of superconformal theory
corresponding to $R^4$ (in the light cone gauge) and $\Delta$
is the contribution to $L_0$ from the descendants of the minimal
models and from  $R^4$ times 28 free fermion theory. The GSO condition
on the right movers is the same as (\ref{gso}), while for the left
movers it is:
\be
\frac{m-m'}{N}-\frac{s+s'}{2}-\frac{\vec{\delta}
\cdot{\vec{w}}}{4}\in
2\bf{Z},
\label{gsob}
\ee
where now $\vec{\delta}$ is the spinor weight of $SO(28)$ and $\vec{w}$ is
the $SO(28)$ weight that the state carries. In addition there is also
the conditions analogous to (\ref{same}) which ensure all the components
are either in the same sector (either R or NS). This is the standard GSO
projection that gives rise to $Spin(32)/Z_2$ theory in the flat case.

Recall that $m'$ and $\bar{m}'$ are the left and right moving momenta
of the free scalar that appears in the $SL(2)/U(1)$ coset. If these
quantities are conserved then they would denote charges with respect
to a $U(1)_L\times U(1)_R$ symmetry. However, as has been discussed in
\cite{gk1,gk2}, only the momenta $(m'+\bar{m}')/2$ are conserved while the
windings $(\bar{m}-\bar{m}')/2$ are conserved only modulo $N$. This is due
to the fact that this CFT is perturbed by an operator that carries $N$
units of windings. Besides this
symmetry associated with the isometry group of the internal space, we
have also the left moving $SO(28)$ Kac-Moody algebra coming from the
28 free fermions as well as the $SU(2)_f$ current algebra which is part
of the left moving ${\cal{N}}=4$ superconformal algebra. Note that, unlike the
$SU(2)$
current algebra in the right moving supersymmetric sector which is
broken by the picture changing operator, the $SU(2)_f$ is a good
symmetry of the theory. The $U(1)_f$ subalgebra of
the $SU(2)_f$ is part of the ${\cal{N}}=2$ subalgebra of the ${\cal{N}}=4$
superconformal algebra, which acts on the individual minimal models
(the remaining generators mix the two minimal models). Important point
to note is that $U(1)_L$ and $U(1)_f$ are not orthogonal to each other.
Since the question of orthogonality involves only the $SL(2)/U(1)$
theory,
it is sufficient to look at the ${\cal{N}}=2$ superconformal algebra for
this system. This system can be represented in terms of a free scalar
$Y$,
a Feigin-Fuchs field $\varphi$ and one complex fermion $\psi_{\pm}$
\cite{es}. The
algebra is given by the following generators:
\begin{eqnarray}
T&=& -\frac{1}{2} (\partial Y)^2 -\frac{1}{2}(\partial \varphi)^2 -Q
\partial^2 \varphi - \frac{1}{2}({\psi_+}\partial \psi_- -
\partial \psi_+ \psi_-)  + T', \nonumber\\
G_{\pm}&=& (\partial \varphi \pm \partial Y) \psi_{\pm} + Q \partial
\psi_{\pm} + G'_{\pm} , \nonumber\\
J&=& \psi_+ \psi_- + iQ\partial Y + J'
\label{algebra}
\end{eqnarray}
where $Q$ is the Feigin-Fuchs background charge and is equal to
$\sqrt{2/N}$.
$T'$, $G'_{\pm}$ and $J'$ are the contributions from the $SU(2)/U(1)$ theory.
Now the vertex operator for a state that carries $m'$ quantum number,
is $\exp{im'QY/2}$, therefore $J_L = i\frac{2}{Q}\partial Y$ measures
the $U(1)_L$ charge $m'$. The singular part of the relevant OPE's are
\be
J_L(z) J(w) = \frac{2}{(z-w)^2} ~~~~, ~~~~ J(z) J(w) = \frac{2}{(z-w)^2},
\ee
where in the second equation we have used the fact that the
${\cal{N}}=2$
algebra implies that the coefficient is given by the $\hat{c}$ of the
full system (namely $SL(2)/U(1) \times SU(2)/U(1)$) which is 2. This
impies that the current $\tilde{J}_L$ which is orthogonal to $U(1)_f$
current $J$ and therefore to the current algebra of $SU(2)_f$ is
\be
\tilde{J}_L= J_L - J.
\ee
$\tilde{J}_L$ is the generator of $\tilde{U}(1)_L$ which is orthogonal
to the $SU(2)_f$. This will be important for us in the following.
For later purpose let us also recall that by bosonizing $U(1)_f$
current $J$, the raising and lowering
generators $J_{\pm}$ of $SU(2)_f$ can be expressed as exponentials of the
corresponding boson. In fact these are just the squares of the spin
fields that take one from NS to R sector. In particular the states in a given
representation
of $SU(2)_f$ can be obtained from a particular state by spectral flow.

The states that give rise to poles in their two-point functions, and
hence couple to the world volume theory, are the ones for which
$|m'-s'|\geq
\ell'+2$ and $|\bar{m}'-\bar{s'}| \geq \ell' +2$.
We will restrict ourselves to such
states. The massless bosonic states (i.e. coming from the right moving
NS sector)  among them which satisfy the GSO conditions
are characterized by $\ell +
\ell'=N-2$, $m=\epsilon \ell$, $m'=-\epsilon(\ell'+2)$,
$\bar{m}={\tilde{\epsilon}} \ell$ and
$\bar{m}'=-{\tilde{\epsilon}}(\ell'+2)$,
where $\epsilon$ and $\tilde{\epsilon}$ are $\pm 1$ and refer to the
possible independent choices of chiral or anti-chiral primaries on the
right and left sectors. Furthermore in the left-moving sector, in
order for them to satisfy the Virasoro and GSO conditions, they must
have $(s, s', \vec{w})$ equal to $(2,0,0)$ or $(0,2,0)$ or
$(0,0,\vec{v})$, where $\vec{v}$ are the weights of $SO(28)$ vector
representation. For the states $(0,0,\vec{v})$ with $m=\epsilon \ell$
and $m'=-\epsilon(\ell'+2)$ we must also include the states obtained
by spectral flow mentioned above in order to construct the full
$SU(2)_f$ representation. Under this flow these states go to
$(2,2,\vec{v})$ with $m=\epsilon(\ell+2)$ and $m'=-\epsilon \ell'$.

The $s=2$ and $s'=2$ states are the $G_{\pm}$ descendants of the $s=0$
and $s'=0$ states. Since the states above are chiral (anti-chiral)
primaries, $G_+$ ($G_-$) will annihilate them. While $m=+\ell$ and
$m=-\ell$ are chiral and anti-chiral primaries respectively, the
states $m'= \ell'+2$ and $m' =-(\ell'+2)$ are chiral and anti-chiral
respectively. This can be seen by using the fact that the Vertex
operator for the
primary $(\ell',m')$ is given by $\exp(-(\ell'+2)Q\phi + i m' QY)$ and
applying $G_{\pm}$ given in (\ref{algebra}) on it. The $U(1)_f$
quantum numbers of these states can then be simply computed by adding
the quantum numbers of the primary and of the $G_{\pm}$.
We can now summarize various quantum numbers in table 1.

\begin{table}
\begin{tabular}{lcllc}
${\bf (s,s',\vec{w})}$~~~~~~~     &${\bf \ell}$~~~~~~~~~~
&${\bf  U(1)_R}$~~~~~~~~~~~
&${\bf  \tilde{U}(1)_L}$~~~~~~~~~~~      &${\bf SU(2)_f}$  \\
$(2,0,0)$ &$\ell=1,..,(N-2)$ &$\tilde{\epsilon}(N-\ell)/2$
&$\epsilon
(N-\ell)/2$  &${\bf 1}$\\
$(0,2,0)$        &$\ell = 0,..,(N-2)$
&$\tilde{\epsilon}(N-\ell)/2$   &$\epsilon (N-\ell)/2$ &${\bf 1}$\\
$(0,0,\vec{v})$    &$\ell = 0,..,(N-2)$
&$\tilde{\epsilon}(N-\ell)/2$
&$\epsilon (N-\ell-1)/2$       &${\bf 2}$
\end{tabular}
\caption{\it Quantum Numbers of Massless states
for $SO(32)$ heterotic theory}
\label{table1}
\end{table}

Note that in the first row $\ell=0$ does not appear since for this
case we have the ground state of the $SU(2)/U(1)$ CFT which does not
have a $G'_{\pm}$ descendant. Starting from each of the above chiral (
antichiral) states, by the spectral flow in the left moving
supersymmetric sector we get all the states that fill out a
hypermultiplet.
Thus we have altogether $(2N-3)$ $SO(28)\times SU(2)_f$
singlet hypermultiplets and $(N-1)$ hypermultiplets transforming as
$(28,2)$ under the flavour group.

As mentioned before $U(1)_R$ and $\tilde{U}(1)_L$ are not separately
conserved due to the presence of the cosmological constant in the
Liouville CFT. While half the difference (say $p$) is conserved, half
of the sum ($w$) is
conserved modulo N. This would suggest that the symmetry group is
$U(1)\times Z_N$. From the above table, however we note that the
the states in the third row which are $SU(2)_f$ doublets, have half
integer values of $p$ and $w$. Therefore the symmetry group is
$\tilde{U}(1)\times Z_{2N}\times SU(2)_f \times Spin(28)/(Z_2)^3$,
where $\tilde{U}(1)$ is the double cover of $U(1)$, and the modding by
$(Z_2)^3$ keeps only 4 conjugacy classes
$({\rm{even}},{\rm{even}},1,sc)$,
$({\rm{even}},{\rm{even}},2,sp')$,
$({\rm{odd}},{\rm{odd}},2,v)$ and $({\rm{odd}},{\rm{odd}},1,sp)$ out
of the total of 32 classes. We will see the appearance of double
coverings
of $U(1)$ and $Z_N$ in the following when we discuss the heterotic
background corresponding to this CFT.

Before proceeding further, let us discuss the case of $E_8\times E_8$
heterotic string. The discussion is exactly as before with the
exception that in the left moving GSO condition (\ref{gsob}),
$\vec{\delta}$ and $\vec{w}$ are the $SO(12)$ spinor weight and the
weight carried by the state respectively. The remaining $SO(16)$
fermions have an independent spin structure sum which gives rise in
the usual way to the unbroken $E_8$ group. The massless states are
again given by the table 1, with $\vec{v}$ denoting now the $SO(12)$
vector. There are however, now additional masssless states that come
from the left moving R sector. It is easy to verify that these states
are given by $(s,s',\vec{w})=(1,1,\vec{sp})$ and
$(m,m')=(\ell+1,\ell'+1)$ where $\vec{sp}$ are the weights of the
spinor representation of $SO(12)$. The $U(1)_f$ quantum number of such states
is zero. The complete table for the massless states together with
various quantum numbers is given in table 2.
\begin{table}
\begin{tabular}{lcllc}
${\bf (s,s',\vec{w})}$~~~~~~~     &${\bf\ell}$~~~~~~~~~~~~~    &${\bf
U(1)_R}$
~~~~~~~~~~~~ &${\bf  \tilde{U}(1)_L}$~~~~~~~~~~~        &${\bf SU(2)_f}$  \\
$(2,0,0)$    &$\ell=1,..,(N-2)$   &$\tilde{\epsilon}(N-\ell)/2$      &$\epsilon
(N-\ell)/2$  &${\bf 1}$\\
$(0,2,0)$        &$\ell = 0,..,(N-2)$
&$\tilde{\epsilon}(N-\ell)/2$  &$\epsilon (N-\ell)/2$   &${\bf 1}$\\
$(0,0,\vec{v})$    &$\ell = 0,..,(N-2)$    &$\tilde{\epsilon}(N-\ell)/2$
&$\epsilon (N-\ell-1)/2$       &${\bf 2}$\\
$(1,1,\vec{sp})$  &$\ell=0,..,(N-2)$ &$\tilde{\epsilon}(N-\ell)/2$
&$\epsilon (N-\ell-1)/2$       &${\bf 1}$
\end{tabular}
\caption{\it  Quantum Numbers of Massless states
for $E_8\times E_8$ heterotic theory}
\label{table2}
\end{table}

The states in the last two columns transform under $SO(12)\times
SU(2)_f$
as $(12,2)+(32,1)$ representations which together form the $(56)$-
dimensional
representation of $E_7$. This is as expected since we know that the
left-right symmetric CFT (corresponding to identification of spin
connection and the gauge connection) breaks one of the $E_8$'s to
$E_7$. Exactly as in the $SO(32)$ case, here also the symmetry is
$\tilde{U}(1)\times Z_{2N}\times E_7/(Z_2)^2$, where the modding by
$(Z_2)^2$
keeps only 2 conjugacy classes $({\rm{even}},{\rm{even}},1)$ and
$({\rm{odd}},{\rm{odd}},56)$ out of the total of 8 classes.

\noindent{\bf Heterotic backgrounds corresponding to the heterotic CFT's}

Now let us turn to the question of what heterotic backgrounds do
these CFT's describe. We will first discuss the heterotic $SO(32)$
theory since in this case the one can use the 5-brane world volume
theory in the S-dual D5 branes of Type I theory. The brane
world volume theory carries symplectic gauge group $Sp(N)$ together
with one hypermultiplet $Y$ transforming in anti-symmetric representation of
$Sp(N)$
and 32 hypermultiplets $q$ in fundamental
representations which transform as vector
representation of $SO(32)$ flavour group. Under the transverse
$SO(4)=SU(2)_R
\times SU(2)_L$, $Y$  transforms in $(2,2)$ and $q$ in $(2,1)$
representations. Furthermorte $q$'s satisfy a reality condition
\be
q*_{I \alpha}= \epsilon_{\alpha \beta} \Omega_{IJ} q_{J \beta}
\label{reality}
\ee
where
$\alpha, \beta =1,2$ denote the $SU(2)_R$ indices, $I,J = 1,..,2N$
denote
$Sp(N)$ indices and $\Omega$ is the symplectic two-form. It is
convenient to write the $2N$ indices $I$ as pair of indices
$(\dot{\alpha},i)$, where $\dot{\alpha}=1,2$ and $i=1,..,N$. The
symplectic form $\Omega_{(\dot{\alpha} i) (\dot{\beta} j)}=
\epsilon_{\dot{\alpha} \dot{\beta}} \delta_{ij}$ and the reality
condition on $q$ just says that it consists of $N$ quaternions $q_i$ with
$SU(2)_R$
acting on the right and $i$-th $SU(2)$ subgroup in the decomposition
$Sp(N)
\rightarrow (SU(2))^N$ acting on the left.

We will now break the symmetries in two steps. Firstly we give a large
vev (of order $\rho$) to the fundamentals to break $Sp(N)$ to
$Sp(N-1)$. This will
correspond to giving a large scale to one of the $N$ instantons as in
\cite{gk}. The $SO(32)$ symmetry is broken down to $SO(28)\times
SU(2)_f$.
The states that become massive at this step can then be
ignored,
since in the limits we will be interested in they will be essentially
infinitely massive. Furthermore massless states (and we will have some
of them) that are neutral under the remaining $Sp(N-1)$ will not be
localized on the remaining $N-1$ branes. These will be the zero modes
associated with the single instanton with large scale. Indeed in the
large scale limit, the corresponding zero modes will be spread all
over the transverse directions. We can therefore also ignore such
states,
since they would decouple from the world-volume physics in the double
scaling limit.

In the second step, we will further break the $Sp(N-1)$ completely at
a scale $\lambda << \rho$, and take a double scaling limit $\lambda
\rightarrow 0$ keeping $\lambda/g_{st}$ finite. This will be done in
such a way that $SU(2)_R\times SU(2)_L$ global symmetry is broken to
$U(1) \times Z_N$.

It is convenient to express the D-terms of ${\cal{N}}=1$ in 6-dim. in
the language of 4-dimensional ${\cal{N}}=1$ where they appear as F-
and
D-terms. Denoting by $A$ and $B$ the two chiral fields contained in
$Y$
(so that $A$ and $B$ have weights $(1/2,1/2)$ and $(1/2,-1/2)$ with
respect
to $SU(2)_L\times SU(2)_R$ respectively), the F-terms are
\be
F =  [A,B] + {\vec{q}} \sigma^{+}{\vec{q}}^{\dagger}
\ee
and the D-terms are
\be
D= [A,\bar{A}] -[B,\bar{B}] + {\vec{q}}\sigma_3 {\vec{q}}^{\dagger}
\ee
where the vectorial notation refers to the $SO(32)$ vector and a dot
product with respect to $SO(32)$ vector is understood. In fact we will
be considering these vectors to lie in a four dimensional subspace
corresponding to an $SO(4)$ subgroup of $SO(32)$. The most general
solution (upto gauge equivalences)
to the above equations, which has $U(1) \times Z_N$ symmetry referred
to above, is
\begin{eqnarray}
B&=&\lambda {\rm diag}(1,\omega, \omega^2, .., \omega^{N-1})~~, ~~~~
A=0  \nonumber\\
{\vec{q}}_i &=& \rho \vec{y} \equiv \rho {\vec{\sigma}} ~~~ , ~~~~
\vec{\sigma}=({\bf{1}},
i\sigma^1,  i\sigma^2, i\sigma^3)
\end{eqnarray}
where $\omega= e^{2\pi i/N}$. The four $y$'s appearing
above are just unit quaternions that are orthogonal to each
other. This was the solution given in \cite{witten}. In fact for the above
solution the contribution of $q$, $A$ and $B$ to the F- and D-terms
above separately vanish. Let us analyse the symmetries of this
solution. For $\lambda=0$, the gauge symmetry is broken to $Sp(N-1)$
as can be readily seen by making a gauge transformation so that
\be
\vec{q} \rightarrow \vec{q}' ~~~ , ~~~ \vec{q}'_i = \delta_{i N}
\sqrt{N}\rho
\vec{y}
\label{basis1}
\ee
Such a gauge transformation can be done by an element $G$ of $Sp(N)$
which (expressed as $N\times N$ matrices with quaternion entries) is
\be
G_{ij} = \frac{1}{\sqrt{N}}
g^{(i-1)(j-1)} ~, ~~~~~g^N={\bf{1}} , ~~~~~g\in SU(2)
\label{basis2}
\ee
On the other hand for $\rho=0$, the $N$ branes are sparated on a
circle so that the gauge symmetry is broken down to $SU(2)^N$. When
both
$\rho$ and $\lambda$ are non-zero $Sp(N)$ is completely broken. As for
the flavour symmetry, $SO(28)$ is clearly unbroken since $\vec{q}$ is
only
along 4-directions. The remaining $SO(4)$ acts on the 4 unit
orthogonal quaternions $\vec{y}$ by rotating them. An $SU(2)$ subgroup
of this $SO(4)$ (which we denote by $SU(2)_f$) can be however un-done
by the diagonal gauge $SU(2)$ subgroup of $SU(2)^N \in Sp(N)$ and the
remaining
$SU(2)$ by the $SU(2)_R$. Thus only the diagonal subgroup of the last
two $SU(2)$'s survives and forms the new $SU(2)_R$ symmetry (in the
absence of $\lambda$). For $\lambda \neq 0$, $SU(2)_L\times SU(2)_R$
is further broken down to $U(1)\times Z_N$ where the $U(1)$ acts on
$A$ (i.e. it is $U(1)_{L+R}$) while the $Z_N$ rotation $B$ can be un-done
by a $Sp(N)$ transformations which $Z_N$ cyclic permutations of the
eigenvalues of $B$ (note that this acts trivially on $q$).
Thus the symmetry group is $U(1)\times Z_N \times SU(2)_f \times
SO(28)$.

Since $\rho >> \lambda$, it is instructive to analyze the light
spectrum by first taking $\lambda=0$. As mentioned above the unbroken gauge
group at this stage is $Sp(N-1)$. The massless hypermultiplet spectrum
can be easily obtained by going to the basis (\ref{basis1},
\ref{basis2}) and
expanding the fields around the vev in the F- and D-terms. The $Sp(N)$
anti-symmetric fields
$Y$
remain massless and they decompose under the unbroken $Sp(N-1)$ as
an anti-symmetric field $Z$, a singlet and 2 fundamental fields $Q_{\alpha}$
which transform as a doublet of $SU(2)_f$. The $Q$'s satisfy a reality
condition analogous to (\ref{reality}) with $SU(2)_R$ replaced by
$SU(2)_f$. There are in fact 4 copies of $Z$ and $Q_{\alpha}$
transforming as $(2,2)$ of $SU(2)_L\times SU(2)_R$. We can carry out a
similar analysis for the
fundamental $Sp(N)$ fields. The result is summarized in
table 3.
\begin{table}
\begin{tabular}{lllc}
${\bf Fields}$ ~~~~~~~~    &${\bf Sp(N-1)}$~~~~~~~~~~
&${\bf  SU(2)_R\times SU(2)_L}$~~~~~~~~~~~
&${\bf  SU(2)_f \times SO(28)}$   \\
$Z$ &$(N-1)(2N-3)-1$ &$(2,2)$      &$(1,1)$   \\
$Q$       &$2(N-1)$  &$(2,2)$
&$(2,1)$ \\
$q$    &$2(N-1)$    &$(2,1)$
&$(1,28)$ \\
$Z'$ &$1$ &$(2,2)$
&$(1,1)$ \\
$q'$  &$1$ &$(3,1)+(1,1)$ &$(1,1)$ \\
$q''$ &$1$ &$(2,1)$ &$(2,28)$
\end{tabular}
\caption{\it Quantum Numbers of Massless scalar fields
after breaking $Sp(N)$ to $Sp(N-1)$}
\label{table3}
\end{table}
Here the symmetries are the new modified symmetries that leave the vev
invariant. $Z'$ and $q'$ are the moduli (position, gauge orientation
and the scale) of the large scale single
instanton. These fields are clearly not localized in the world
volume. $q''$ are $SO(28)$ vectors but are also not localized in the
world volume \footnote{Note that in a symmetric comactification of heterotic
theory, the total number of massless $SO(28)$ vectors is topological
and does not change when we go to the limit of small scale
instantons.}. Therefore the fields $Z'$, $q'$ and $q''$ will decouple
from the world volume physics. In the above table $Q$ transforms under
$SU(2)_L\times SU(2)_f$ as $(2,2)$ representation. This was because it originally was in the
antisymmetric representation $Y$ of $Sp(N)$. However in $\rho
\rightarrow
\infty$ limit where we can restrict ourselves to $Sp(N-1)$ theory
there is an enhancement of symmetry. The action is invariant when $Q$
is
transformed by $SU(2)'\times SU(2)_f$ where $SU(2)'$ is independent of
$SU(2)_L$. Putting together $Q$ and $q$ therefore we get an $SO(32)$
vector
representation. This is what was used in the analysis of \cite{gk}.
Indeed in the corresponding CFT, the left moving sector had $SU(2)_L
\times
SO(32)$ symmetry.

The second step of symmetry breaking (i.e. $\lambda \neq 0$),
corresponds to turning on $Z$ and $Q$. As mentioned above $Sp(N-1)$
is completely broken and $SU(2)_L \times SU(2)_R$ is broken to $U(1)
\times
Z_N$ where the $U(1)$ acts on the $A$ direction and $Z_N$ on the $B$
direction. There are $(2N-3)$ complex massless scalars coming from $Z$
and $Q$ with $U(1)\times Z_N$ charge $(1,0)$ and 2 complex scalars
each with $(0,m)$ with $m=2,3,..,(N-1)$ and one with charge $(0,0)$.
These are all $SU(2)_f \times
SO(28)$ singlets. Finally from $q$ we have one complex scalar each
with
charge $(m/2,m/2)$ for $m=1,..,(N-1)$ which transform as $(2,28)$
under $SU(2)_f \times SO(32)$.
The appearance of half integer charges implies extension of $U(1)$ and
$Z_N$
to $\tilde{U}(1)$ and $Z_{2N}$ with certain $Z_2$
identifications  as implied by the
correlation of charges. Since this analysis only involves massless
states,
we do not see any $SO(28)$ spinors. However, the symmetry group
$\tilde{U}(1)\times Z_{2N}\times SU(2)_f\times Spin(28)/(Z_2)^3$ is
certainly a symmetry of the above spectrum.

The total number of massless scalars therefore agrees with the
counting coming from CFT analysis given in table 1. However, as it
happens in the type II case, the charge assignments are
different. This is because the CFT states should couple to the gauge
invariant composite operators in the world volume theory. We can
construct $Sp(N-1)$ invariant composite operators from the fields
$Z$, $Q$ and $q$ exactly as in the type II case. Since $Z$ and $Q$
transform as $(2,2)$ under $SU(2)_R\times SU(2)_L$, it is convenient
to
denote $Z_A$, $Q_A$ along the $A$ direction and $Z_B$ and $Q_B$ along
the $B$ direction. Gauge invariant chiral composite operators together
with their quantum numbers are given in the table 4.
\begin{table}
\begin{tabular}{lcllc}
{\bf Operators}~~~~ &$m$~~~~~~~~~~~~~~~~~
&${\bf U(1)_R}$~~~~~~~~~~ &${\bf  U(1)_L}$
~~~~~~~~~~~&${\bf  SU(2)_f \times SO(28)}$   \\
${\rm tr} Z_A^m$ &$m=2,..,N-1$ &$m/2$ &$m/2$  &$(1,1)$     \\
${\rm tr} Z_B^m$ &$m=2,..,N-1$ &$m/2$ &$ -m/2$  &$(1,1)$ \\
$Q_A^t Z_A^m Q_A$ &$m=0,..,N-2$ &$(m+2)/2$  &$(m+2)/2$ &$(1,1)$ \\
$Q_B^t Z_B^m Q_B$ &$m=0,..,N-2$ &$(m+2)/2$ &$-(m+2)/2$ &$(1,1)$ \\
$Q_A^t Z_A^m q$ &$m=0,..,N-2$ &$(m+2)/2$ &$(m+1)/2$   &$(2,28)$\\
$Q_B^t Z_B^m q$ &$m=0,..,N-2$ &$(m+2)/2$ &$-(m+1)/2$   &$(2,28)$
\end{tabular}
\caption{\it Quantum Numbers of chiral composite operators
of $Sp(N-1)$ theory}
\label{table4}
\end{table}
Comparing this with table 1 (for $\tilde{\epsilon}=+1$ which
corresponds to the chiral primaries), we find complete agreement.

It is worth noting that although the above discussion made use of the
effective world volume theory (namely $Sp(N)$ gauge theory) in the
S-dual D5 brane, we could have come to the same conclusions, directly,
by working with the zero modes around the heterotic theory solution.
The latter approach will have the advantage that it would not be tied
to the details of the 5-brane world volume theory and therefore will
be applicable also to the $E_8\times E_8$ heterotic theory.

The $SO(32)$ heterotic theory solution corresponds to $N$-instantons in an
$SU(2)$ subgroup of
the $SO(32)$ gauge theory in the target space. One of these $N$
instantons is of large size, while the others are approximately point
like.
The zero modes of interest are the ones localized near the point-like
instantons, since they are the ones that will appear in the world
volume physics. Imbedding $SU(2)$ in $SU(2)\times SU(2)_f\times
SO(28)$ maximal subgroup of $SO(32)$, the zero modes can be described
in terms of ADHM construction of the $SO(4)$ instantons. The $SO(32)$
gauge fields then transform as $(3,1,1)+(1,3,1)+ (1,1,{\rm Ad})+
(2,2,28)$ under the maximal subgroup. The instanton involves
non-trivial
background only in the $(3,1,1)$ piece. Apart from the usual zero
modes associated with the $SU(2)$ instantons, the  zero modes will
also
come from the $(2,2,28)$ piece. The instanton solution and the
zero modes can be conveniently described in terms of ADHM data. There
is however a huge redundancy in this data which is given by the action of
the ADHM symmetry group. It is therefore natural to describe the zero
modes in terms of the ADHM group invariant quantities. The ADHM
construction of the $SU(2)$ instantons can be done in different
ways. For example, imbedding $SU(2)$ in $SO(4)=SU(2)\times SU(2)_f$,
we can think of it as $SO(4)$ instanton. The ADHM group in this case
is $Sp(N)$ and in fact the world volume analysis given above is precisely
this case; the D terms are just the ADHM constraints. On the other
hand we can also describe the $SU(2)$ instantons directly with the
ADHM group being $SU(N)$ (or $SO(N)$ if we view $SU(2)$ as $Sp(1)$).
It turns out however that $SU(N)$ description does not yield the
zero modes associated with the gauge fields in $(2,2,28)$
representation
since the latter satisfy a reality condition. $SU(N)$ is broken down
to $SO(N)$ due to this reality condition.
These different descriptions (i.e. $Sp(N)$ or $SO(N)$) must provide
the same physical
information of the zero modes if we restrict to the ADHM group invariant
quantities. Let us therefore obtain  the above results (which
corresponded to the ADHM group being  $Sp(N)$) by working instead with $SO(N)$
ADHM group.

We will now be general and consider some gauge group $G$ which has a maximal
subgroup
$Sp(1)\times G'$ such that adjoint representation of $G$ splits into
adjoints of $Sp(1)$ and $G'$ plus $(2,R)$ where $R$ is a pseudo-real
representation of $G'$. The $(2,R)$ representation further satisfies a
reality condition
\be
A_{\mu}^*= \sigma_2 A_{\mu}\Omega
\label{real}
\ee
where $A_{\mu}$
is a $(2\times {\rm dim}(R))$ matrix and $\Omega$ is a symplectic matrix
associated with $R$.
For $G=SO(32)$, $G'=SU(2)_f\times SO(28)$ with
$R=(2,28)$ and for $G=E_8$, $G'=E_7$ with $R=(56)$.

The ADHM data for N instantons in $Sp(1)$ gauge theory is given in
terms  of a $(1+N)\times N$ matrix $\Delta$ with quaternion entries
\be
\Delta_{\lambda, i } = a_{\lambda, i } +
b_{\lambda, i} x; ~~~~  x = x_{\mu} \sigma^{\mu};
\ee
where $\lambda=1,..,2+2N$, $i,j=1,..,N$ and $a$ and $b$ are
constants. By using the symmetries one can
choose $b$ to be (writing $\lambda=u+ j \alpha$ with $u=1$ being the
$Sp(1)$ gauge group index)
\be
b_{u,i}=0; ~~~~ b_{j, i} = \delta_{ji} {\bf 1}
\ee
where ${\bf 1}$ is the unit element of the quaternion. It is
convenient to define
\be
a_{u,i}= w_{i}; ~~~~a_{j,i} = a'_{j,i}= a'^{\mu}_{ji}
\sigma^{\mu}
\ee
$a'^{\mu}$ transform in the symmetric tensor representation of the ADHM
symmetry group $SO(N)$, while $w$ transforms as bi-fundamental of the
gauge group $Sp(1)$ and $SO(N)$. The left and right $SU(2)$ actions on the
quaternion entries in $a'$
are respectively the $SU(2)_L$ and $SU(2)_R$ actions ($SU(2)_L\times
SU(2)_R$
is the Euclidean group acting on the 4-dim. space on which the
instantons live and therefore is the transverse group to the 5-brane
world volume). In other words the $\mu$ index in $a'$ transforms as a
vector of the $SO(4)=SU(2)_L\times SU(2)_R$. On the other hand the
left and right actions on the quaternions in $w$ are respectively the
$Sp(1)$ gauge group and $SU(2)_R$ actions respectively. $\Delta$
satisfies a quadratic constraint
\be
({\bar{\Delta}} \Delta)_{ij }=f^{-1}_{ij} {\bf
1}
\ee
where $f$ is a real symmetric non-singular $N\times N$ matrix and
transforms in the symmetric tensor representation of $SO(N)$.
This results in quadratic constraints on $a'$ and $w$
in the adjoint of $SU(2)_R \times SO(N)$
which are just the D-terms for the $SO(N)$ theory. The instanton
solution
for the $Sp(1)$ adjoint gauge field is
\be
A_{\mu} = \bar{U} \partial_{\mu} U ;~~~~
\bar{U}\Delta=\bar{\Delta}U=0,~~~
\bar{U}U={\bf 1}
\ee
where $U$ is a $(N+1)\times 1$ matrix with quaternion entries.
Solution (upto symmetries) to these
constraints that breaks $SU(2)_L\times SU(2)_R$ to $U(1)\times Z_N$ is
\be
w_{i}=\rho {\bf 1} ~~~{\rm for~ each}~~ i
\ee
and $a'^{\mu}$ being diagonal  matrices with $N$ different eigenvalues
that are $Z_N$ symmetric in a plane. In other words, writing
$a'^1+ia'^2=A$
and $a'^0+ia'^3 =B$, this corresponds to choosing $A=0$ and $B=\lambda {\rm
diag}(1,\omega, \omega^2,..,\omega^{N-1})$ with $\omega=e^{2\pi
i/N}$. As in the previous discussion we
take $\rho$ to be large while $\lambda$ small. The ADHM group then is
broken to an approximate $SO(N-1)$ group. $A$ ($B$) splits into
symmetric tensor
$Z_A$ ($Z_B$) and a fundamental $Q_A$ ($Q_B$) of
$SO(N-1)$.

Finally we have to also consider the zero modes coming from the gauge
fields in $(2,R)$ representation of $SU(2)\times G'$. These can be
explicitely given as
\be
A_{\mu}^{(2,R)} = {\bar{U}}_{\lambda}\partial_{\mu} \Delta_{\lambda, i}
f_{ij} q_{j}
\ee
where $q_j$ are constant $2\times {\rm dim}(R)$ matrices satisfying a
reality condition
\be
q^*_j= \sigma_2 q_j \Omega
\ee
$q$ therefore transforms as $(N,2,R)$ under
$SO(N)\times SU(2)_R \times G'$ together with the reality condition.
Note that with this reality condition on $q$, the gauge fields
$A_{\mu}^{(2,R)}$ satisfies the required reality condition (\ref{real}).
In the limit of large $\rho$ only
the $SO(N-1)$ vector remains localized zero mode (the remaining
one spreads over the entire transverse space and therefore is irrelevant to
the world-volume physics). We will, by a slight abuse of notation, in
the following denote by $q$ the localized zero modes in $SO(N-1)$ vector
representation.

Note that for $G'=SU(2)_f\times SO(28)$,
$Z$, $Q$ and $q$ obtained here differ from the ones preceding the
table 3, in that here $Q$ is $SU(2)_f$ singlet instead of doublet and
$q$ vice versa. This will, however, not matter when we construct the ADHM
group invariant quantities. The chiral ones among these are
$tr Z_A^m$ and $tr Z_B^m$ for $m=2,..,N-1$ and $Q_A^T Z_A^m Q_A$
and $Q_B^T Z_B^m Q_B$  for
$m=0,..,N-2$ which result
in the quantum numbers of the first 4 rows of operators in table 4.
Besides this, we have also operators $Q_A^T Z_A^m q^+$ and
$Q_B^T Z_B^m q^+$ for $m=0,..,N-2$, with $q^+$ denoting the highest
weight of $SU(2)_R$. The quantum numbers of these operators
reproduce the last two rows in table 4.
Thus we see that the precise details of the world-volume theory,
namely
whether it is $Sp(N)$ or $SO(N)$ gauge theory was not necessary for
obtaining the matching with the CFT. In fact it is determined by the
zero mode structure around the instanton solution.

We can now apply this analysis to the case of $E_8$, since it does not
require the details of the 5-brane world volume theory (which is not
known at present). It is easy to see that
for $G=E_8$ and $G'=E_7$ with the representation $R=(56)$ the
ADHM group invariant combinations given in the previous paragraph
reproduce the table 2 coming from the CFT analysis. Alternatively one
might have also worked with the $SO(4)$ instanton giving rise to the
ADHM group $Sp(N)$, and go through the above steps to construct
$Sp(N)$ invariant quantities which also gives the result of Table 2.
Thus although the matching with the chiral states of the CFT does not
tell us much about the details of the 5-brane world-volume theory, it
does suggest that it should have a local $Sp(N)$ or $SO(N)$ symmetry.
How this symmetry is realized in a $(1,0)$ theory of tensor and
hypermultiplets is an important open question.

\section{Conclusions and Open Questions}

In this paper, we have studied the physics of NS 5-branes in heterotic
like theories with 16 supercharges. Non-singular CFT describing the
string
theory in these backgrounds is given by an ${\cal N}=2$ minimal model
together with and $SL(2)/U(1)$ Kazama Suzuki model. We have considered
two classes of models; one which is obtained by projecting type II theories
by $(-1)^{F_L}\sigma$ with $\sigma$ being a half shift on a circle
longitudinal to the 5-brane. In this case the S-duality relation (for
the IIB case), which
maps the NS 5-brane to D5 brane in the S-dual model, is
expected to hold due to adiabatic argument. We indeed find that the
CFTs
describe holographically the chiral gauge invariant operators in the
dual theory. Furthermore we constructed the boundary states
corresponding to massive charged states, which are non-BPS in the
present case, and showed that their charges and tensions agree with
the results expected from the brane world volume theory. It is
interesting
to note that although we are here dealing with non-BPS states their
tensions are not renormalized.

The second type of models we considered are the standard $SO(32)$ and
$E_8\times E_8$ heterotic theories. Our discussion was restricted to
the case of symmetric 5-branes, which gives rise to ${\cal N}=(2,2)$
CFT (actually this symmetry is extended to $(4,4)$). In this case
the internal CFT turns out to be the same as above coset models
(modulo the left moving gauge fermions). The $SU(2)_R\times SU(2)_L$
symmetry
is broken down to $U(1)\times Z_N$, while the gauge symmetry is broken
down to $SU(2)\times SO(28)$ and $E_7\times E_8$ respectively. We obtained the
chiral states which among others included states transforming as
$(2,28)$ and $(56,1)$ under the above groups.  For the $SO(32)$ case,
by studying the D-terms of the $Sp(N)$ gauge theory living in the brane
world volume, we
identified the ground state which has the above symmetry and breaks
completely the $Sp(N)$ group. We found
that the gauge invariant chiral operators match with the CFT states.
In the 10-dimensional heterotic theory this ground state corresponds
to $N$ instantons in an $SU(2)$ subgroup of $SO(32)$, with one of the
instantons having large scale while the remaining $N-1$ are separated
and have small scales. We studied the ADHM construction of such
instanton configurations and showed that both $Sp(N)$ and $SO(N)$
ADHM groups give the same spectrum of chiral ADHM group invariant
operators. This allowed us to verify the holographic correspondence
also for the $E_8\times E_8$ theory where we do not know the 5-brane
world volume theory.

Although in this paper we have discussed 5-branes in the 10-dimensional
supersymmetric heterotic theories, the discussion can be easily
generalized to the non-supersymmetric heterotic theories. For example
the tachyon free $SO(16)\times SO(16)$ model is obtained from $SO(32)$
theory by a $Z_2$ orbifolding generated by the element $(-1)^F
\sigma_s$ where $\sigma_s$ is the shift by the weight $(0,sp)$ in the
$SO(16)\times SO(16)$ decomposition of $SO(32)$. We can obtain the
crresponding CFT by starting from the one given here and modding out
by this $Z_2$ orbifold group.

There are several open questions. The most important one is an
understanding of the massive charged states in the heterotic theory.
In the first type of models we have analyzed, they appeared as
D-strings (or D2 branes) stretched between the separated NS5 branes
and
we had a boundary state description for them. In the heterotic theory,
on the hand there are no D-branes and boundary states. Therefore
the description of these excitations remains an interesting open
problem.

In the heterotic theory we considered only the symmetric
5-branes. There
are moduli in the CFT which are charged under $SU(2)\times SO(28)$
or $E_7$ in the two theories. By turning on these moduli we can go
to a non-symmetric situation cooresponding to a $(4,0)$ CFT. This
amounts to having the $N$ instantons not in an $SU(2)$ subgroup of
$SO(32)$ or $E_8$. However in the $E_8\times E_8$ case, these
deformations still keep all the $N$ instantons in one $E_8$ factor.
It will be interesting to construct CFTs which describe backgrounds
with $N_1$ instantons in one $E_8$ and $N-N_1$ instantons in the
second $E_8$. This will be also necessary in order to construct CFTs
for CHL models\cite{chl}.

{\bf{Acknowledgements}} K.S.N. would like to thank the participants
to the 2001 Chrete Regional Meeting in String Theory,
especially O. Aharony and D. Kutasov, for useful discussions.

% {\bf{Appendix}}

%\newpage


\begin{thebibliography}{99}

\bibitem{seiberg} N. Seiberg, Phys. Lett. {\bf 408B} (1997) 98.

\bibitem{abks} O. Aharony, M. Berkooz, D. Kutasov and
N. Seiberg, JHEP {\bf 9810} (1998) 004.

\bibitem{gk1} A. Giveon and D. Kutasov, JHEP {\bf 9910} (1999) 034.

\bibitem{gk2} A. Giveon and D. Kutasov, JHEP {\bf 0001} (2000) 023.

\bibitem{gk} M. Gremm and A. Kapustin, JHEP {\bf 9911} (1999) 018.

\bibitem{ds} D. Diaconescu and J. Gomis, Nucl. Phys. {\bf B548} (1999) 258.



\bibitem{adg} I. Antoniadis, S. Dimopouls and A. Giveon,
JHEP {\bf 0105} (2001) 055.

%\bibitem{bianchi} M. Bianchi and Y. Stanev, Nucl.Phys. {\bf B523} (1998) 193.

\bibitem{chs} C. Callan, J. Harvey and A. Strominger, {\it Lectures
at the 1991 Trieste Spring School on String Theory}, hep-th/9112030.

\bibitem{kpr} C. Kounnas, B. Rostand and M. Porrati, Phys. Lett.
{\bf 258B} (1991) 61.

\bibitem{afk} I. Antoniadis, S. Ferrara and C. Kounnas, Nucl. Phys.
{\bf B421} (1994) 343.

\bibitem{kounnas} C. Kounnas, hep-th/0012192.

\bibitem{ADHM} M. Atiyah, V. Drinfeld, N. Hitchin and Yu. Manin,
Phys. Lett. {\bf 55A} (1978) 185.

\bibitem{es} T. Eguchi and Y. Sugawara, Nucl. Phys. {\bf B598} (2001) 467.

\bibitem{hk} K. Hori and A. Kapustin, hep-th/0104202.

\bibitem{vm} S. Mukhi and C. Vafa, Nucl. Phys. {\bf 407B} (1993) 667.

\bibitem{gv} D. Ghoshal and C. Vafa, Nucl. Phys. {\bf 453B} (1995) 121.

\bibitem{ov} H. Ooguri and C. Vafa, Nucl. Phys. {\bf 463B} (1996) 55.

\bibitem{yamaguchi} S. Yamaguchi, hep-th/0102176.

\bibitem{sagnotti} G. Pradisi, A. Sagnotti and  Y.S. Stanev,
Phys. Lett. {\bf B381} (1996) 97.
C. Angelantonj, M. Bianchi, G. Pradisi, A. Sagnotti and
Ya.S. Stanev, Phys.Lett. {\bf B387} (1996) 743.

\bibitem{rs} A. Recknagel and V. Schomerus, Nucl. Phys. {\bf 531B} (1998) 185.

\bibitem{douglas} I. Brunner, M.R. Douglas, A. Lawrence
and C. Romelsberger, JHEP {\bf 0008} (2000) 015.

\bibitem{vafa} K. Hori, A. Iqbal and C. Vafa, hep-th/0005247.

\bibitem{lerche} W. Lerche, hep-th/0006100.

\bibitem{lls} W. Lerche, A. Lutken and C. Schweigert, hep-th/0006247.

\bibitem{bgmn} M. Bianchi. E. Gava, J. F. Morales and K.S. Narain,
Nucl. Phys. {\bf B547} (1999) 96.

\bibitem{ghmn1} E. Gava, A.B. Hammou, J. F. Morales and K.S. Narain,
Nucl. Phys. {\bf B605} (2001) 17.

\bibitem{ghmn2} E. Gava, A.B. Hammou, J. F. Morales and K.S. Narain,
JHEP {\bf 0103} (2001) 035.

\bibitem{gutp} M. Gutperle, JHEP {\bf 0008} (2000) 036.

\bibitem{witten} E. Witten, Nucl. Phys. {\bf B460} (1996) 541.

\bibitem{witten1} E. Witten, {\it Some Comments on String Dynamics},
hep-th/9507121.

\bibitem{chl} S. Chaudhuri, G. Hockney and J.D. Lykken,
Phys.Rev.Lett. 75 (1995) 2264.

\end{thebibliography}
\end{document}